\documentclass[12pt]{iopart}
\newcommand{\ee}{\hbox{e$^+$e$^-$}}
\begin{document}

\title[Neutrino factory R\&D $–$ global perspective]{Neutrino Factory
R\&D $-$ a global perspective}

\author{B. Foster\dag\   
}

\address{\dag\ H.H. Wills Physics Laboratory, University of
Bristol, Bristol, BS8 1TL, UK and\\
DESY, Notkestrasse 85, 22761 Hamburg, Germany
\vskip0.5cm Invited talk at the Nufact02 Workshop, Imperial College of Science, Technology and Medicine, London,
July 2002}

\begin{abstract}
The current world status of plans for future particle-physics accelerators
and accelerator research and development is outlined. The developments for
a neutrino factory are placed in this context. Finally, a specific initiative relevant for Europe is discussed. 
\end{abstract}



\maketitle

\section{The global picture}
There is at the moment an unprecedented world-wide consensus on the next
steps to be taken in the construction of major `energy-frontier' accelerator
facilities. This consensus has been crystallised in documents from all three
major regions active in experimental particle physics: the ACFA report on an \ee\ linear collider~\cite{ACFAdoc}; the ECFA report on the future of accelerator-based particle physics in Europe~\cite{ECFAdoc}; and the HEPAP report on Long-Range Planning for U.S. High-Energy Physics~\cite{HEPAPdoc}.
The first priority for future facilities is the construction, in a timely
manner, of an \ee\ linear collider with centre-of-mass energy of at least around 500 GeV. This consensus was supported by the report of the OECD Global Science Forum 
Consultative Group on High-Energy Physics~\cite{OECDdoc}, whose members are drawn from active particle physicists, science policy makers and representatives of governments and funding agencies. The report states that:

`The Consultative Group concurs with the world-wide consensus of the scientific community that a high-energy electron-positron linear collider is the next facility on the Road Map. There should be a significant period of concurrent running of the CERN Large Hadron Collider (LHC) and the LC, requiring the LC to start operating before 2015. Given the long lead times for decision-making and for construction, consultations among interested countries should begin at a suitably chosen time in the near future.'

\section{The neutrino factory}
While the unanimity of view expressed above is both of the first importance and gratifying, it does not address the subject of this conference, i.e. a neutrino factory. Fortunately, several of these documents expressed views in this area, and once again there was a substantial consensus that, although a linear \ee\
collider was the top priority for the future of particle physics, it was also important to have other priorities. The importance of the currently
approved programme was emphasized, in particular completing the LHC and its detectors, and ensuring that running experiments, for example those at the Tevatron and HERA, were able to exploit their full potential and ensure value for money from the major investment that has been made in them. More directly relevant for this
conference, there was also a clear appreciation of the importance of general accelerator R\&D for the future of our field in general, and for the neutrino factory in particular. The ECFA report explicitly mentions this in the Executive
summary by recommending:
`a co-ordinated collaborative R\&D effort to determine 
the feasibility and practical design of a neutrino factory based on a high-intensity muon storage ring'. The ECFA report also explicitly recognises the importance and promise of the neutrino factory and supports the 
increase in resources and manpower in the field of
accelerator physics necessary to build it:
`A neutrino factory complex, beginning with its proton 
driver, allows a number of unique experiments in a 
fundamental domain: neutrino masses and mixing, CP 
violation, and lepton number violation. The realisation of 
this important programme requires a substantial 
international programme of R\&D.' Similarly positive statements are to be
found in the HEPAP report. Although ACFA does not specifically mention a neutrino factory, the resources devoted to this end in Japan as well as the world leadership in neutrino physics currently enjoyed by our Japanese colleagues is eloquent enough testimony to the importance that
such a development has in Asia. 

The importance of an increase in activity on accelerator R\&D was also recognized by the OECD report. This was part of a general perception that accelerator R\&D was not only important for the future of particle physics, but also for the improvement of existing, and the development of new, 
accelerator-based research facilities vital in many other areas of science.

\section{A specific initiative in Europe}

It is one thing for bodies such as ECFA to produce documents stating that more money spent on particle physics in general and accelerator R\&D in particular is a good thing. It is somewhat more difficult to actually obtain such resources from funding agencies. In order to try to bridge this gap, ECFA is taking a rather concrete initiative in the context of an `Integrated Infrastructure Initiative' in the European Union's Framework VI programme.
The main goals of this integrated infrastructure initiative are:
\begin{itemize}
\item to develop closer interactions between the various players 
in accelerator research and technology;
\item to develop interactions between accelerator physicists, 
universities, potential users and industry;
\item to create a distributed but coherent activity in Europe in the 
domain of accelerator physics and technology (joint research
 activities).
\end{itemize}
The proposed activities include high-intensity proton-accelerating cavities, 
high-power targets, (with spin-off in the domain of spallation sources),  
muon cooling and items aimed towards a neutrino factory. 
The budget request is likely to be approximately 30 MEuro over 5 years, and
is made in conjunction with a funding commitment from
the partners of 120 MEuro. There is also some possibility of applying
for funds to enhance already approved projects that are currently
in the construction stage. The proposal is currently under intensive preparation under the coordination of Roy Aleksan and with Saclay being the lead laboratory. The deadline for submission is likely to be in March 2003, with the call for proposals being sent out before the end of 2002. There is of course extremely strong competition for these funds, including from fields close to ours, such as astroparticle and nuclear physics. Nevertheless, if we were to be successful, accelerator R\&D in Europe would be significantly strengthened.

\section{Conclusion}
Although the main thrust of accelerator R\&D in the next few years will be on an \ee\ linear collider, there is wide recognition of the excellent physics case for the construction of a neutrino factory. There still remain significant accelerator R\&D questions that must be solved before a realistic proposal can be made. It is imperative that the resources currently available for such research should be increased; in order to do so, we should attempt to tap in to funding sources beyond those to which we  normally have access. With the LHC on the horizon, and proposals such as the linear collider and neutrino factory gaining increasing momentum, we are in for an exciting time in the next few years.

\end{document}